\documentclass[aps,prb,twocolumn,groupedaddress,showpacs]{revtex4}
\usepackage{longtable}
\usepackage{graphicx}% Include figure files
\usepackage{dcolumn}% Align table columns on decimal point
\usepackage{bm}% bold math
\newcommand{\PCCO}{Pr$_{2-x}$Ce$_x$CuO$_4$}

\newcommand{\QCP}{quantum critical point}

\newcommand{\FS}{Fermi surface}

\newcommand{\AFM}{antiferromagnetic}
\newcommand{\MR}{magnetoresistance}

\newcommand{\etal}{\emph{et al.}}
\newcommand{\cotet}{$\cot(\theta_H)$}
\begin{document}

\title{Hole superconductivity in the electron-doped superconductor Pr$_{2-x}$Ce$_x$CuO$_4$}


\author{Y. Dagan}
\email[]{yodagan@post.tau.ac.il} \affiliation{School of Physics
and Astronomy, Raymond and Beverly Sackler Faculty of Exact
Sciences, Tel-Aviv University, Tel Aviv, 69978, Israel}
%\affiliation{Center for Superconductivity Research Physics
%Department University of Maryland College Park, MD, 20743}
\author{R.L. Greene}
\affiliation{Center for Superconductivity Research Physics
Department University of Maryland College Park, MD, 20743}


\date{\today}

\begin{abstract}
We measure the resistivity and Hall angle of the electron-doped
superconductor Pr$_{2-x}$Ce$_x$CuO$_4$ as a function of doping and
temperature. The resistivity $\rho_{xx}$ at temperatures $100
K<T<300 K$ is mostly sensitive to the electrons. Its temperature
behavior is doping {\emph{independent}} over a wide doping range
and even for non superconducting samples. On the other hand, the
transverse resistivity $\rho_{xy}$, or the Hall angle $\theta_H$
where $\cot(\theta_H)=\rho_{xx}/\rho_{xy}$, is sensitive to both
holes and electrons. Its temperature dependence is strongly
influenced by doping, and \cotet~ can be used to identify optimum
doping (the maximum T$_c$) even well above the critical
temperature. These results lead to a conclusion that in electron
doped cuprates holes are responsible for the superconductivity.
\end{abstract}

%\keywords{}

\maketitle
\section {Introduction}

A striking property of the high T$_c$ cuprate superconductors is
the extreme sensitivity of many of their electronic properties to
the number of charge carriers put into the copper-oxygen planes
(doping). From a chemical point of view these charge carriers can
be either holes, as in La$_{2-x}$Sr$_x$CuO$_4$, or electrons as in
\PCCO~. However, in the electron-doped superconducting cuprates,
transport \cite{WuJiang,fournier2bands} and angle resolved
photoemission spectroscopy (ARPES) studies
\cite{armitageprldoping} have shown that both electrons and holes
play a role in the normal state properties. An interesting and
important question is whether both also play a role in the
superconducting state \cite{JEHirsch}.
\par
Conventional superconductors are characterized by a single
temperature scale, T$_c$, above which all the information about
their superconducting properties is lost and they become normal
metals. This is not the case for the hole-doped cuprate
superconductors. It is believed that due to strong electron
correlations doping effects on many electronic properties are seen
at relatively high temperatures. For example, by looking at
resistivity curves of various doping levels of one compound well
above T$_c$ one can identify the location of the doping level with
maximum T$_c$ (optimum doping)\cite{Takagi}. A different picture
is seen on the electron-doped side of the phase diagram. Near
optimum doping the temperature dependence of the Hall coefficient
(R$_H$), along with some other transport properties, were
interpreted as evidence for two types of carriers
\cite{WuJiang,fournier2bands}. ARPES measurements indeed revealed
an evolving Fermi surface from small electron pockets at low
dopings to a Fermi surface with holes and electrons like regions
with hot spots at optimum doping
\cite{armitageprldoping,matsui:047005}.
\par
In hole doped cuprates the resistivity is linear in temperature
over a wide temperature range for underdoped samples extrapolating
to zero at T=0 for optimally doped samples and quadratic in
temperature on the overdoped side \cite{Takagi}. The Hall angle
follows a T$^2$ dependence \cite{chienHallangle}. This was
interpreted in the framework of Fermi Liquid theory by the
existence of hot spots, small regions on the Fermi surface with
very short scattering time
\cite{CarringtonHS},\cite{stojkovicandpines},\cite{kontani}, or
"cold spots" \cite{ioffeandmillis},\cite{ZheleznyakYakovenkoDrew}.
N. E. Hussey suggested an anisotropic T$^2$ scattering rate
combined with T independent scattering rate \cite{HusseyEPJB}.
Other non Fermi liquid ideas involved, two different scattering
times for the charge and spin channels,\cite{andersonHallangle} or
the Marginal Fermi Liquid theory with a linear in T, isotropic
scattering rate and a temperature independent small angle impurity
scattering \cite{AbrahamsandVarmaHall}. In overdoped
Bi$_2$Sr$_{2-x}$La$_x$CuO$_6$ \cite{AndoHallanglepower} and in
Bi$_2$Sr$_2$Ca$_{n-1}$Cu$_n$O$_y$
\cite{KonstantinovicHallanglepower} a deviation from the T$^2$
behaviour was observed, the exponent $\alpha$ in the fit
\cotet~$=a+bT^\alpha$ decreased with increasing doping. This
behaviour was interpreted as a contribution of extended regions on
the Fermi surface to the Hall angle as the doping level increases
\cite{HusseyEPJB}.
\par
In the electron doped cuprates, Woods \etal~ \cite{woods} reported
that in optimally doped samples $\alpha$ is twice as large as the
resistivity exponent. They interpreted this behavior in the
framework of the theory of Abrahams and Varma
\cite{AbrahamsandVarmaHall}. A possibility of hole
superconductivity in the electron-doped cuprates was speculated by
Z. Z. Wang \etal~ \cite{WangandOngHoleSC} on the basis of Hall and
resistivity measurements for presumably overdoped samples. W.
Jiang \etal~ \cite{WuJiang} suggested that holes are crucial for
the occurrence of superconductivity in electron-doped
superconductors based on magneto-transport measurements on oxygen
treated Nd$_{1.85}$Ce$_{0.15}$CuO$_{4-\delta}$. Qazilbash \etal~
\cite{qazilbashRamanSCstate} have shown from Raman spectroscopy
measurements that superconductivity in the electron-doped is
primarily due to pairing and condensation of hole-like carriers.
It was also theoretically predicted that superconductivity will be
favored by having hole states rather than electron ones at the
Fermi energy \cite{JEHirsch}.
\par
The detailed doping and temperature study of resistivity and Hall
angle reported here enables us to qualitatively follow the
contributions of electrons and holes to the transport and to
deduce their respective roles in generating the superconducting
condensate.
\par
\section{Samples preparation and measurements.}
\PCCO~ \textit{c}-axis oriented films of various cerium doping
concentrations: $x=0.11, 0.12\ldots0.19$ were deposited from
stoichiometric targets on (100) oriented SrTiO$_3$ substrates
using the pulsed laser deposition technique as described elsewhere
\cite{daganResistivityPRL}. The films were patterned to form Hall
bars using ion milling. The Hall angle was measured at 14T where
all the samples are normal and $\rho_{xy}$ has a linear dependence
on magnetic field. The normal state resistivity and the
superconducting transition temperatures and widths are identical
to the previously reported data\cite{daganResistivityPRL}
\par
\section{Results and discussion.}
\begin{figure}
\includegraphics[width=1\hsize]{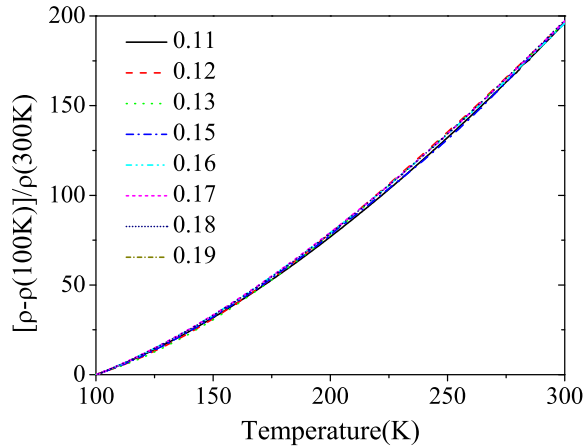}
\caption {Normalized resistivity from 100K to 300K. The
resistivity at 100K is subtracted from $\rho(T)$ then all curves
are normalized at 300K. Except for a residual term and a
coefficient all doping concentrations exhibit the same temperature
dependence.\label{rscaling}}
\end{figure}
First, we show, that in a strong contrast to the hole doped
cuprates, the doping level has no influence on the temperature
dependence of the resistivity, $\rho$ and that the resistivity is
dominated by the electrons. In figure \ref{rscaling} we plot
$[\rho(T)-\rho(100K)]/\rho(300K)$. This merely cancels the
contribution of any residual impurity scattering and divides each
curve by a numerical factor. We chose 100 K for two reasons: a) at
this temperature all the resistivity curves have approximately the
same slope; b) this temperature is still well above the upturn in
the resistivity. Remarkably, all the data collapse on a single
curve. The non superconducting sample $x=0.11$ scales together
with all the other superconducting ones. In this sample there are
no holes as can be inferred from high field Hall measurements,
where $\rho_{xy}$ is found to be linear in H and negative up to 60
T at 100K~ \cite{PengchengLi} as expected for a single type of
carrier. We can therefore conclude that the doping independent
behavior of the resistivity for $x=0.11-0.19$ must be due to the
electrons. We note that such data collapse is not possible for the
hole-doped cuprates where the temperature dependence of the
resistivity changes from linear to quadratic as the doping is
increased.
\par
\begin{figure}
\includegraphics[width=1\hsize]{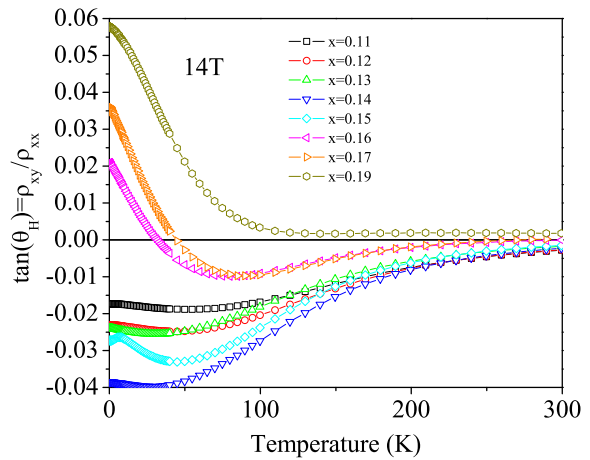}
\caption {$\tan\theta_H=\rho_{xy}/\rho_{xx}$ at 14T as a function
of temperature for the various doping levels. $\tan\theta_H(T)$
has a clear doping dependence.\label{tanthetah}}
\end{figure}
\par

\begin{figure}
\includegraphics[width=1\hsize]{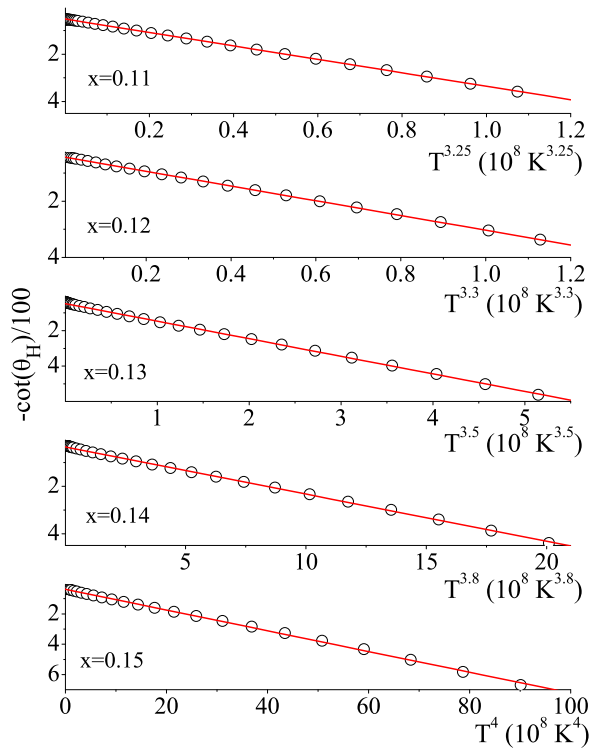}
\caption{$\cot\theta_H$ as a function of $T^\alpha$, where
$\alpha$ is found from a fit to $\cot(\theta_H)=a+bT^\alpha$ for
the temperature range 100-300 K. Note that $\alpha$ increases with
increasing doping. For $x\geq0.16$ \cotet~ cannot be fit to a
power law.}\label{cothetah}
\end{figure}

Second, we show that the Hall angle is sensitive to the doping
level and that optimum doping can be identified using this
property. This result is due to a hole contribution to the
transverse resistivity. In Figure \ref{tanthetah} we show
$\tan(\theta_H)$ at 14T for $0.35K<T<300K$ for
$x=0.11,0.12\ldots0.19$. The Hall angle changes sign with doping
and temperature. This indicates that the transverse resistivity is
sensitive to both holes and electrons. The reported T$^2$
dependence of \cotet~ for hole doped cuprates is not seen here.
Instead $\alpha$, the exponent obtained from the fit to
\cotet~$=a+bT^\alpha$ for $100K<T<300K$, changes with doping. In
Figure \ref{cothetah}, \cotet~ is shown for under-to-optimally
doped \PCCO~, $x=0.11-0.15$, as a function of T$^\alpha$. The
exponent $\alpha$ increases monotonically from $3.24$ for $x=0.11$
to $4$ for $x=0.15$. For the overdoped region $(x\geq0.16)$ the
power law behavior is lost and such a fit is not possible. The
exponent found for $x=0.15$ is consistent with previous reports
\cite{fournier2bands},\cite{woods}. While the resistivity above
T$_c$ can give no indications for the doping level, the doping
level and in particular that of maximum T$_c$ can be identified
using the exponent of \cotet~at least on the under-to-optimum
doping regime $(0.11 \geq x \geq0.15)$ even at relatively high
temperatures as can be clearly seen in Figure \ref{cothetah}.
\par
\par
In hole doped cuprates $\alpha$ decreases when the doping is
increased from optimum to the overdoped region
\cite{AndoHallanglepower},\cite{KonstantinovicHallanglepower}. A
similar, but mirror like, picture is seen here for the optimum to
underdoped region, where $\alpha$ increases with increasing doping
reaching its maximum value at optimum doping (before the power law
behaviour is lost). It was suggested that the difference between
the Hall and resistivity temperature dependences in hole doped
cuprates is coming from the hot-spots on the Fermi surface
\cite{HusseyEPJB}. For electron-doped cuprates ARPES measurements
have found that the hot-spots can be seen most clearly at optimum
doping. Since most of the scattering takes place at these (hole
like) hot spots one expects the largest difference between the
resistivity and the Hall angle exponents at this doping level, as
seen in Figure \ref{cothetah}. At low dopings the Fermi surface
consists only of electron pockets \cite{armitageprldoping}. In
this case electrons should dominate both resistivity and Hall
angle. One should therefore expect the exponents of these two
transport properties to become closer as the doping level is
decreased from optimum, as observed. Summarizing, the behavior of
the Hall angle is consistent with both hole and electron regions
on the \FS~ contributing to the transverse resistivity.
\par
Although the Fermi surface is electron like from $x=0.11$ all the
way up to optimum doping, as inferred from the negative sign of
R$_H$ at all temperatures \cite{daganResistivityPRL}, there is a
strong hole contribution to R$_H$ in optimally doped samples. This
is suggested by the steep rise in R$_H$ towards positive values as
the temperature is decreased below 67 K
\cite{WangandOngHoleSC,daganResistivityPRL}. Above optimum doping
the Fermi surface rearranges and becomes hole like, presumably at
a quantum critical point \cite{daganResistivityPRL}. Away from
this \QCP~ there is a funnel shaped region of quantum and thermal
fluctuations in the doping-temperature phase diagram, resulting in
the reappearance of both the electron and the hole bands at higher
temperatures even for overdoped samples. The phase diagram
presented by Li \etal~ \cite{PengchengLi} from high field Hall and
\MR~ measurements may define these different regions. At low
temperatures $(T<10K)$ on the overdoped side $(x\geq0.17)$,
outside of the funnel shaped region of quantum fluctuations, the
resistivity and \cotet~ follow the same temperature dependence,
thus suggesting a metallic-like single band Fermi surface.
Additional evidence for the dominance of a single band at low
temperatures is found from thermopower and Hall measurements
\cite{li_thermopower, daganResistivityPRL}. These two transport
properties yield exactly the same carrier concentrations at low
temperatures when analyzed using simple single band Drude mode. We
also note that for the overdoped side as the Ce concentration is
increased the number of holes and T$_c$ decrease. The reason for
the vanishing of T$_c$ in both types of cuprate is yet to be
understood.
\par

The origin of the resistivity behavior at high temperatures is
unclear at the moment. Its doping independence suggests that it is
unrelated to the \AFM~ order or to the hot spots in the Fermi
surface. Hublina and Rice \cite{HlubinaRice} showed that cold
regions can short out the effect of the hot spots on the \FS~.
This results in a resistivity which is insensitive to the hot
spots (and doping). In our case not only is the resistivity (above
100K) insensitive to the hot spots but also to the development of
the hole like regions on the Fermi surface. This dominance of the
electrons needs further theoretical investigation. While the
electron-doped resistivity is very different from that of the
hole-doped cuprates there is some resemblance in the behavior of
the Hall resistivity (or the Hall angle) for the two types of
cuprates. First, it has a doping dependence even at high
temperatures and optimum doping can be identified using the power
of the temperature dependence of \cotet~. Second, a strong hole
contribution to R$_H$ appears at optimum doping. This leads us to
conclude that holes play a similar role in both types of
superconductor. The absence of a hole contribution in the
underdoped, non-superconducting, samples and the lack of a doping
dependence for the electron-dominated resistivity, strongly
suggest that electrons have no (or a very small) contribution to
superconductivity in the electron-doped cuprate superconductors.
\section{\label{summarySec}Summary}
We measured the resistivity and the Hall angle of \PCCO~ as a
function of temperature and doping from  $x=0.11$ (underdoped and
nonsuperconducting) to $x=0.19$ (very overdoped). While the
temperature dependence of the resistivity between 100K and 300K
show no variation with doping, the exponent $\alpha$ of the Hall
angle in the fit $\cot(\theta_H)=a+bT^\alpha$ exhibits doping
dependence. This quantity can be correlated with the occurrence of
superconductivity. We have shown that the resistivity is mostly
sensitive to the electrons while the transverse resistivity probes
both the hole and electron regions on the Fermi surface. Our
results lead us to conclude that in electron-doped cuprates holes
are responsible for superconductivity.

\begin{acknowledgments}
We thank Guy Deutscher and Girsh Blumberg for very useful
discussions. Support from NSF grant number DMR-0352735 is
acknowledged for work at the University of Maryland. Y.D. wishes
to thank the German Israeli foundation for support.
\end{acknowledgments}

\bibliographystyle{apsrev}
\bibliography{holeSC}
\end{document}